\documentclass[a4paper]{article}
\usepackage{ISCSLP2026}

\usepackage{multirow}
\usepackage{tabularx}
\usepackage{placeins}
\usepackage{enumitem}
\usepackage{url}
\usepackage[hidelinks]{hyperref}

\newcommand{\train}{\mathcal{D}}
\newcommand{\Ls}{\mathcal{L}}
\newcommand{\E}{\mathbb{E}}

\title{
Long-Tail Rebalancing for Non-Verbal Vocalization-Aware ASR:
\protect\\
A Track~1 System for the NVVSpeech Challenge
}

\name{
    Shangyue Jia$^1$, Jingru Ma$^2$, Yangzhuo Li$^2$, Daoping Luo$^3$, Bowen Tian$^2$, Hanchen Lu$^2$ \\
    Wenze Ren$^4$, Yunxiang Chen$^5$, Houdun Liu$^5$, Su Feng$^5$, Lei Xie$^6$, Liumeng Xue$^{1,7}$\sthanks{$^*$Corresponding author.}
}

\address{
    $^1$School of Intelligence Science and Technology, Nanjing University \\
    $^2$Jilin University; $^3$Beijing Institute of Technology \\
    $^4$National Taiwan University; $^5$Shenzhen Pimei Technology Co., Ltd. \\
    $^6$Audio, Speech and Language Processing Group (ASLP@NPU), School of Computer Science, \\
    Northwestern Polytechnical University \\
    $^7$State Key Laboratory of Novel Software Technology, Nanjing University
}

\email{
    jason.yue.2026@gmail.com, lmxue@nju.edu.cn
}

\begin{document}

\maketitle
\begin{abstract}
Non-verbal vocalizations (NVVs) carry important paralinguistic information but
are often omitted by conventional automatic speech recognition (ASR) systems.
The ISCSLP NVVSpeech Challenge requires joint transcription of lexical content
and 16 NVV categories under limited and highly imbalanced supervision. We
present a data-centric NVV-aware ASR pipeline based on cross-dataset label
harmonization and a two-stage sampling schedule. We map heterogeneous source
labels to the official taxonomy and exclude samples without a reliable mapping.
Our schedule first uses square-root category sampling to moderate
the long-tailed distribution and then applies uniform-category fine-tuning. On
a fixed local validation split, square-root category sampling performs best among the tested
single-stage settings. The final two-stage system obtains an official score of
63.86 and ranks fourth in Track~1.
\end{abstract}

\section{Introduction}
Non-verbal vocalizations (NVVs), such as laughter, coughing, breathing,
and crying, convey information about speaker state, affect, and conversational
dynamics beyond lexical content~\cite{schuller2013computational,schuller2013paralinguistics,
trouvain2012comparing,li2026nvsynthesis,hu2026lalmpreference}.
Conventional automatic speech recognition (ASR) systems, however, primarily
focus on lexical transcription and often omit these events. Consequently,
lexical-only transcripts fail to capture how speakers feel, react, and
interact.

Recent NVV-annotated datasets have expanded the coverage of non-verbal speech
events through scripted collection, human annotation, and automatic labeling
of real-world speech~\cite{liao2025nvspeech,ye2025nonverbalspeech,
wu2025smiip,mai2026mnv17,cao2026nvvlocator}.
However, these datasets differ in event inventories, annotation
granularity, and data provenance, making joint training across datasets
nontrivial. In parallel, event-aware ASR systems have
begun to represent NVV events as inline output tokens, enabling lexical content
and non-verbal events to be recognized within a unified decoding
sequence~\cite{inaguma2018end,shione2023automatic,yang2026wesr,cao2026sourceadaptive}.
The ISCSLP NVVSpeech Challenge provides a concrete testbed for this problem by
requiring systems to jointly transcribe speech and 16 predefined NVV
categories~\cite{nvvspeechchallenge2026}. Nevertheless, normalizing heterogeneous
labels alone does not resolve the severely long-tailed category distribution:
simply combining all available data leaves rare NVV categories underrepresented.
At the other extreme, aggressively flattening the category distribution does
not necessarily improve the aggregate score. This sensitivity motivates the
question of how heterogeneous NVV resources should be normalized and
rebalanced without distorting the joint lexical--NVV prediction task.

To address this problem, we collect several open-source datasets, map their NVV
labels to the 16 challenge categories, and exclude samples without a reliable
mapping. We then fine-tune Qwen3-ASR with a two-stage schedule. Stage~1 uses
square-root category sampling to smooth the global distribution, and Stage~2
continues training with uniform-category sampling to place greater emphasis on
tail categories.

We evaluate sampling strategies on a fixed local validation split using the
official Track~1 evaluator. Square-root sampling gives the highest local
validation score among the tested single-stage settings, whereas stronger
rebalancing is not monotonically better. The two-stage system gives our highest
official submission score, but the single-stage square-root model scores higher
on the local validation split.

Our contributions are threefold:
\begin{itemize}[
leftmargin=*,
topsep=1pt,
itemsep=0pt,
parsep=0pt,
partopsep=0pt
]
\item We harmonize heterogeneous NVV annotations into the 16-category
challenge taxonomy and filter unreliable mappings for joint lexical--NVV
modeling.

\item We systematically study category-level rebalancing strength under a fixed
backbone, objective, and output format, showing that moderate square-root
sampling outperforms both natural and fully uniform sampling on our local
validation split.

\item We provide complementary evaluation across local validation, the official
challenge test, and a held-out-speaker MNV-17 test set, and our final two-stage
system achieves an official score of 63.86, ranking fourth in Track~1.

\end{itemize}

\begin{figure*}[!t]
    \centering
    \includegraphics[width=\textwidth]{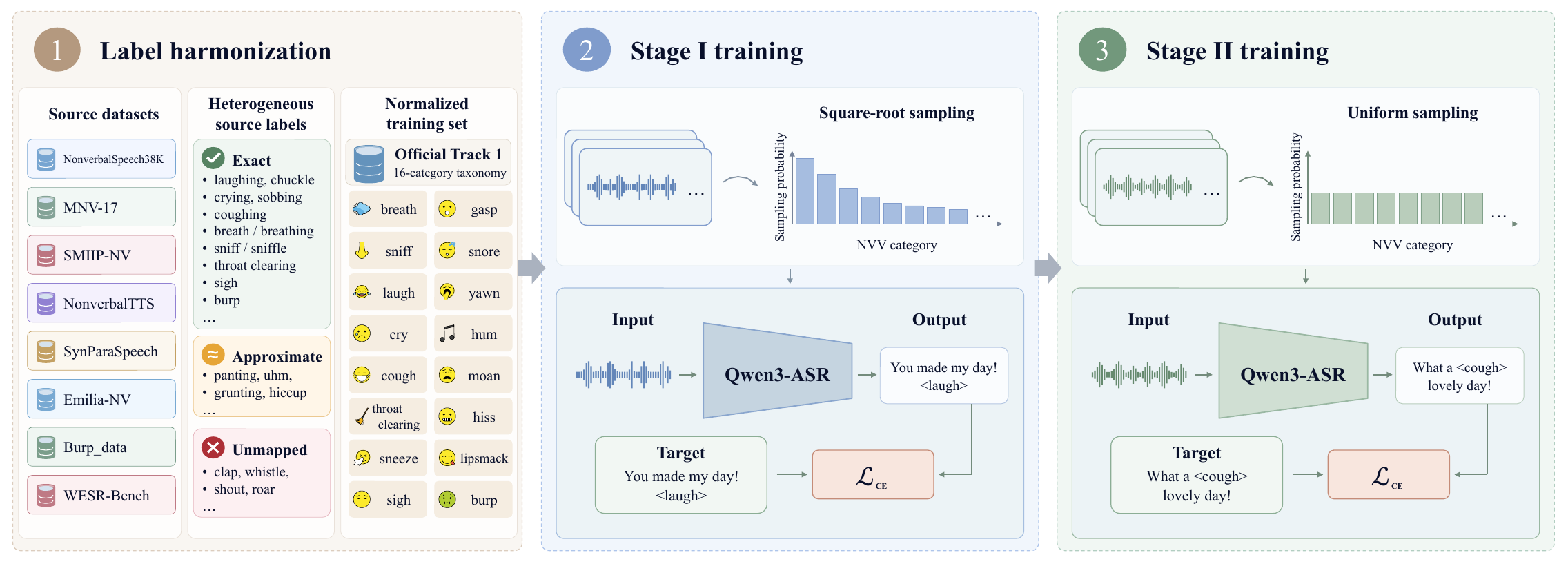}
    \caption{Overview of our data-centric NVV-aware ASR pipeline, consisting of cross-dataset label harmonization, Stage I training with square-root category sampling, and Stage II fine-tuning with uniform-category sampling.}
    \label{fig:label_harmonization}
\end{figure*}

\section{Related work}

\subsection{Event-aware speech recognition}

Most ASR systems prioritize lexical transcription, whereas event-aware systems
also include social or non-verbal events in the output sequence. End-to-end
studies have jointly modeled speech recognition and social-signal detection and
later integrated lexical and non-verbal information in unified recognition
frameworks~\cite{inaguma2018end,shione2023automatic}. More recent work combines
NVV data construction, recognition, and generation by treating NVVs as
decodable elements~\cite{liao2025nvspeech,ye2025nonverbalspeech,li2026wavbench}.
WESR-Bench further defines explicit event categories and position-aware
evaluation~\cite{yang2026wesr}. In contrast, we keep the backbone and output
protocol fixed and study training-data composition and category sampling.

Data-centric strategies such as two-stage training and class balancing can
improve rare-event recognition~\cite{yang2026beyondwords}. However, the setting
in which multiple heterogeneous datasets must first be mapped into a challenge
taxonomy remains less explored. We address this setting without an additional
event detector or augmentation model.

\subsection{Non-verbal vocalization datasets and label harmonization}

NVV datasets differ in event inventories, annotation granularity, languages,
and collection procedures. NonverbalSpeech38K provides large-scale data for
non-verbal speech understanding and generation~\cite{ye2025nonverbalspeech};
MNV-17 focuses on performative Mandarin events~\cite{mai2026mnv17}; and SMIIP-NV
provides multiply annotated expressive Mandarin speech~\cite{wu2025smiip}.
The challenge and WESR-Bench highlight the need for a unified label space and a
consistent evaluation protocol~\cite{yang2026wesr,nvvspeechchallenge2026}.
Direct merging can conflate related but non-equivalent labels and amplify
imbalance because the datasets differ in scale and category coverage. We
therefore map direct and manually verified correspondences to the Track~1
taxonomy and exclude labels without a reliable mapping.

\subsection{Class imbalance and data rebalancing}

Long-tailed recognition work has compared instance-balanced, class-balanced,
square-root, and progressively balanced sampling. Square-root sampling provides
an intermediate distribution, whereas progressive balancing changes the
distribution during training~\cite{kang2020decoupling}. Dynamic Curriculum
Learning jointly adjusts sampling and loss schedules~\cite{wang2019dynamiccurriculum},
and Balanced Meta-Softmax estimates class sampling rates from meta-validation
data~\cite{ren2020balancedmetasoftmax}. We instead keep the objective fixed and
use a simple two-stage schedule: square-root category sampling followed by
uniform-category fine-tuning.
 
\section{Method}

\subsection{Problem formulation}

We formulate NVV-aware ASR as conditional sequence generation. For an audio
waveform $x$ and language identifier $\ell$, the model produces one sequence
$y$ that interleaves lexical words with inline NVV tags. Let
$\train=\{(x_i,\ell_i,y_i)\}_{i=1}^{N}$ denote the processed training set,
and let $q$ denote the example distribution induced by the sampler. The 
ASR model is trained with the standard autoregressive supervised fine-tuning
(SFT) objective
\begin{equation}
\Ls_{\mathrm{SFT}}(\theta;q)
=-\E_{i\sim q}
\left[\log p_{\theta}(y_i\mid x_i,\ell_i)\right],
\end{equation}
while keeping the backbone, tokenizer, target serialization, and loss fixed.
We modify only $q$ through category-aware sampling. The method combines
cross-dataset label harmonization, power-based category sampling, and a
two-stage full-parameter SFT schedule.

\subsection{Cross-dataset label harmonization}

Merging NVV datasets directly is problematic because the same event may appear
under different names, while one source label may be broader or less reliable
than an official Track~1 category. At the same time, dropping every source
label that is not textually identical would discard useful supervision. We
therefore use the official 16-category Track~1 label set as a common target
space and define a fixed mapping from source labels into it, as illustrated
in the label-harmonization stage of \autoref{fig:label_harmonization}.

Specifically, for a source label \(s\), we map it to the corresponding official category when the correspondence is exact or a manually verified approximation; labels without a reliable interpretation are left unmapped and excluded from the harmonized training set. For utterances containing multiple events, we normalize each event label independently while retaining the utterance as a single training example, thereby preserving the complete lexical–NVV sequence and the original event order. The resulting harmonized training set is summarized in Table~\ref{tab:training_data}.

\begin{table}[t]
\centering
\caption{Composition of the normalized NVV-aware ASR training data.}
\label{tab:training_data}
\scriptsize
\setlength{\tabcolsep}{2.5pt}
\renewcommand{\arraystretch}{1.0}

\resizebox{0.95\columnwidth}{!}{%
\begin{tabularx}{\columnwidth}{
@{}
>{\centering\arraybackslash}X
>{\centering\arraybackslash}p{0.15\columnwidth}
>{\centering\arraybackslash}p{0.22\columnwidth}
>{\centering\arraybackslash}p{0.18\columnwidth}
@{}
}
\toprule
\textbf{Dataset}
& \textbf{Lang.}
& \textbf{Samples}
& \textbf{Hours} \\
\midrule

NonverbalSpeech38K~\cite{ye2025nonverbalspeech}
& ZH / EN
& 9,597
& 31.88 \\

MNV-17~\cite{mai2026mnv17}
& ZH
& 1,497
& 4.26 \\

Burp\_data\footnotemark[1]~\cite{nvvspeechchallenge2026}
& ZH / EN
& 1,425
& 2.11 \\

NonverbalTTS~\cite{borisov2025nonverbaltts}
& EN
& 3,518
& 9.14 \\

SMIIP-NV~\cite{wu2025smiip}
& ZH
& 3,008
& 6.06 \\

SynParaSpeech~\cite{bai2026synparaspeech}
& ZH
& 4,000
& 5.51 \\

Emilia-NV~\cite{liao2026emilianv}
& ZH
& 3,011
& 10.02 \\

WESR-Bench~\cite{yang2026wesr}
& EN
& 592
& 1.50 \\

\midrule
\textbf{Total}
& \textbf{ZH / EN}
& \textbf{26,648}
& \textbf{70.48} \\

\bottomrule
\end{tabularx}
}%
\end{table}

\begin{table*}[!t]
    \centering
    \caption{Main results on the local validation split using the official
    Track~1 evaluator. We compare the released Whisper baseline with Qwen3-ASR
    1.7B trained using natural sampling, square-root sampling (SQRT), and the
    two-stage SQRT + uniform schedule. The 10-epoch SQRT setting uses the same
    total number of training epochs as the two-stage system. We report the
    overall validation score, along with separate scores and evaluation metrics
    for Chinese and English.}
    \label{tab:local_validation}
    \setlength{\tabcolsep}{3.2pt}
    \renewcommand{\arraystretch}{1.08}
    \resizebox{\textwidth}{!}{
    \begin{tabular}{
        ccc
        ccccc
        ccccc
    }
        \toprule
        \multirow{2}{*}{\textbf{System}}
        & \multirow{2}{*}{\textbf{Setting}}
        & \multirow{2}{*}{\textbf{Validation score} $\uparrow$}
        & \multicolumn{5}{c}{\textbf{Chinese}}
        & \multicolumn{5}{c}{\textbf{English}} \\
        \cmidrule(lr){4-8}
        \cmidrule(lr){9-13}
        &
        &
        & \textbf{Score} $\uparrow$
        & $\mathbf{F_1}$ $\uparrow$
        & \textbf{mNTD} $\downarrow$
        & \textbf{Err.} $\downarrow$
        & \textbf{CER} $\downarrow$
        & \textbf{Score} $\uparrow$
        & $\mathbf{F_1}$ $\uparrow$
        & \textbf{mNTD} $\downarrow$
        & \textbf{Err.} $\downarrow$
        & \textbf{WER} $\downarrow$ \\
        \midrule
        Official baseline
        & Released Whisper
        & 37.26
        & 48.68 & 0.43 & 0.49 & 0.19 & 13.67\%
        & 25.84 & 0.20 & 0.81 & 0.21 & 11.37\% \\
        Qwen3-ASR 1.7B
        & Natural ($\alpha=1$), 5 epochs
        & 62.33
        & 67.83 & 0.66 & 0.38 & 0.06 & 3.60\%
        & 56.83 & 0.53 & 0.47 & 0.11 & 6.31\% \\
        Qwen3-ASR 1.7B
        & SQRT ($\alpha=0.5$), 5 epochs
        & \textbf{66.25}
        & \textbf{71.06} & \textbf{0.69} & \textbf{0.32} & \textbf{0.06} & \textbf{3.39\%}
        & \textbf{61.45} & \textbf{0.59} & \textbf{0.43} & \textbf{0.11} & 7.10\% \\
        Qwen3-ASR 1.7B
        & SQRT + uniform, 5 + 5 epochs
        & 58.91
        & 65.52 & 0.62 & 0.35 & 0.06 & 3.79\%
        & 52.29 & 0.48 & 0.52 & 0.11 & 6.95\% \\
        Qwen3-ASR 1.7B
        & SQRT ($\alpha=0.5$), 10 epochs
        & 62.72
        & 68.31 & 0.66 & 0.36 & 0.06 & 3.57\%
        & 57.12 & 0.54 & 0.47 & 0.11 & \textbf{6.16\%} \\
        \bottomrule
    \end{tabular}
    }
\end{table*}

\subsection{Power-based category sampling}

\begin{figure}[!t]
    \centering
    \includegraphics[width=0.80\columnwidth]{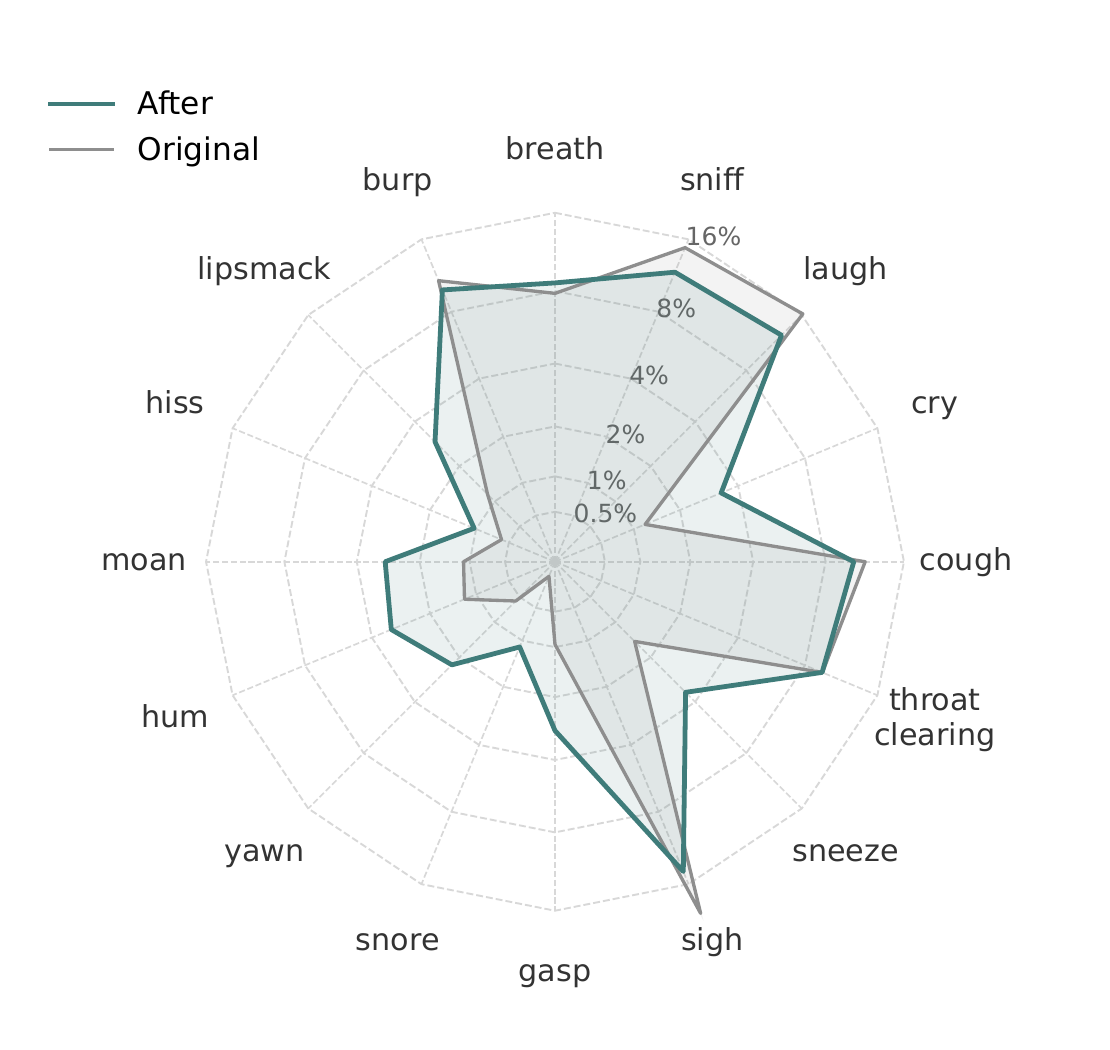}
    \caption{
        NVV-category occurrence shares in the original training set and after
        square-root resampling. The radial axis uses a logarithmic transformation
        to better visualize rare-category exposure.
    }
    \label{fig:category_share_shift}
\end{figure}

As shown in Figure~\ref{fig:category_share_shift}, the category distribution
remains long-tailed even after harmonization.
Natural sampling, which follows the empirical frequencies of the assigned
categories ($\alpha=1$), gives the model many updates on frequent events but
few opportunities to learn the acoustic cues and corresponding NVV labels of
rare events. Inspired by class-level rebalancing
practice~\cite{kang2020decoupling},
we address this mismatch using category-level sampling while leaving each
utterance intact.

Let $c_i$ denote the sampling category assigned to utterance $i$, and let
$n_c$ denote the number of utterances assigned to category $c$. For utterances
containing multiple valid NVV events, we use the first valid NVV label in
target order as the sampling key, while keeping the complete lexical--NVV
target sequence unchanged. Only 1,413 of 26,648 training utterances (5.30\%)
contain multiple distinct valid NVV labels, suggesting that this assignment
has a limited effect on the overall rebalancing distribution. Let
$\mathcal{C}_{\mathrm{train}}=\{c\mid n_c>0\}$ denote the represented
categories. For $\alpha\in[0,1]$, we define
\begin{equation}
p_c(\alpha)=\frac{n_c^{\alpha}}
{\sum_{k\in\mathcal{C}_{\mathrm{train}}}n_k^{\alpha}},
\end{equation}
At each sampling step, we first sample $c\sim p_c(\alpha)$ and then uniformly
sample an utterance from the corresponding category bucket. Hence, the induced
utterance-level sampling probability is
\begin{equation}
q_{\alpha}(i)=\frac{p_{c_i}(\alpha)}{n_{c_i}}.
\end{equation}
When $\alpha=1$, $q_{\alpha}$ reduces to uniform sampling over training
utterances; $\alpha=0.5$ gives square-root sampling
(SQRT)~\cite{conneau2021xlsr}, while $\alpha=0$ assigns equal probability to
each represented category.

\footnotetext[1]{\url{https://huggingface.co/datasets/NVVSpeech-Challenge/Burp_data}}

\subsection{Two-stage sampling schedule}

As illustrated in \autoref{fig:label_harmonization}, our final schedule
consists of square-root sampling in Stage~1 followed by uniform-category
sampling in Stage~2. In \textbf{Stage~1}, we
fine-tune Qwen3-ASR~1.7B with full-parameter updates while moderately increasing
the exposure of tail categories. In \textbf{Stage~2}, we resume from the
Stage~1 checkpoint and continue SFT with equal category probabilities. The
backbone, objective, target serialization, and lexical--NVV sequence format
remain unchanged across stages.

\section{Experiments}

\subsection{Experimental setup}

\noindent\textbf{Backbone selection.} Qwen3-ASR provides a multilingual ASR
backbone with strong Chinese and English recognition capabilities~\cite{shi2026qwen3asr},
making it suitable for the challenge test set, which contains approximately
balanced numbers of Chinese and English utterances. Of the two released variants, 0.6B and 1.7B,
we use the 1.7B model for full-parameter SFT.

\noindent\textbf{Training and inference.} We train with bfloat16 precision, a
global batch size of 64, a learning rate of $2\times10^{-5}$ with linear decay,
a warmup ratio of 0.02, and a gradient clipping threshold of 1.0. In our main
two-stage system, Stage~1 and Stage~2 each run for 5 epochs. For inference, we
use the released Qwen3-ASR implementation with greedy decoding. Local validation
outputs are scored using the official Track~1 evaluator.

\noindent\textbf{Evaluation data.} We evaluate our systems on a local validation
split, the official challenge test set through submission, and a subset of the
public MNV-17 test split. For local model selection, we use a fixed validation
split of 214 utterances: 147 Chinese and 67 English utterances covering all 16
categories. These comprise 108, 86, and 20 utterances from NonverbalSpeech38K,
MNV-17, and Burp\_data, respectively. The MNV-17 test subset is used to assess
generalization to unseen speakers, as detailed in Section~\ref{subsec:mnv17}.

\noindent\textbf{Evaluation metrics.} Following the Track~1
protocol~\cite{nvvspeechchallenge2026}, we use character error rate (CER) and
word error rate (WER) to measure Chinese and English lexical transcription
errors, respectively. We also report the NVV recognition $F_1$ score,
multi-event normalized tag distance (mNTD), and tagged-transcript error (Err.).
On MNV-17, we follow the original benchmark~\cite{mai2026mnv17} and report
joint CER and NVV accuracy, with the latter requiring an exact match between
the complete predicted and reference NVV sequences.

\subsection{Main results and analysis} 

Table~\ref{tab:local_validation} and Table~\ref{tab:official_results} summarize the results on our local validation set and the official challenge test set, respectively. The Stage 1 SQRT system achieves the best local validation score of 66.25 and obtains an official score of 52.61 in the Preliminary Stage. In contrast, the Stage 2 model achieves a lower local validation score of 58.91, but obtains our best official result of 63.86 in the Final Stage.

\begin{table}[!t]
    \centering
    \caption{Official Track~1 baseline and submission scores.}
    \label{tab:official_results}
    \footnotesize
    \setlength{\tabcolsep}{2.5pt}
    \renewcommand{\arraystretch}{1.02}
    \begin{tabularx}{\linewidth}{@{}
        >{\centering\arraybackslash}X
        >{\centering\arraybackslash}p{0.27\linewidth}
        >{\centering\arraybackslash}p{0.20\linewidth}
        @{}}
        \toprule
        \textbf{System} & \textbf{Challenge phase} & \textbf{Official score} \\
        \midrule
        Official Whisper baseline & Final & 33.32 \\
        Qwen3-ASR 1.7B, SQRT & Preliminary & 52.61 \\
        Qwen3-ASR 1.7B, two-stage & Final & \textbf{63.86} \\
        \bottomrule
    \end{tabularx}
\end{table}

Table~\ref{tab:per_category_validation} traces how per-category recognition changes from the SQRT checkpoint to the two-stage checkpoint. Yawn shows the clearest improvement, with $F_1$ moving from 0.09 to 0.39, while throat clearing moves from 0.50 to 0.57. SQRT remains stronger for most other categories, so the two-stage system changes the balance of recognition rather than lifting performance uniformly across the label set.

\begin{table}[!ht]
\centering
\caption{Per-category local-validation results for the SQRT and two-stage
    systems. Val gives the number of reference occurrences for each category;
    $R$ denotes recall, and $\Delta F_1$ denotes the change from SQRT to
    two-stage using the displayed $F_1$ values.}
\label{tab:per_category_validation}
\scriptsize
\setlength{\tabcolsep}{2pt}
\renewcommand{\arraystretch}{0.92}
\begin{tabular}{@{}lrccr@{}}
\toprule
\textbf{Category} & \textbf{Val}
& \textbf{SQRT $F_1/R$} & \textbf{Two-stage $F_1/R$}
& $\boldsymbol{\Delta F_1}$ \\
\midrule
sigh & 20 & 0.69/0.85 & 0.64/0.75 & $-0.05$ \\
laugh & 20 & 0.90/0.95 & 0.82/0.80 & $-0.08$ \\
sniff & 20 & 0.61/0.70 & 0.47/0.45 & $-0.14$ \\
cough & 20 & 0.39/0.35 & 0.26/0.20 & $-0.13$ \\
burp & 20 & 0.95/0.90 & 0.76/0.80 & $-0.19$ \\
throat clearing & 20 & 0.50/0.75 & 0.57/0.80 & $+0.07$ \\
breath & 20 & 0.63/0.65 & 0.32/0.25 & $-0.31$ \\
sneeze & 10 & 0.89/0.80 & 0.86/0.90 & $-0.03$ \\
hum & 11 & 0.96/1.00 & 0.79/1.00 & $-0.17$ \\
cry & 20 & 0.44/0.40 & 0.37/0.25 & $-0.07$ \\
lipsmack & 13 & 0.96/1.00 & 0.80/0.92 & $-0.16$ \\
moan & 11 & 0.91/0.91 & 0.71/1.00 & $-0.20$ \\
gasp & 20 & 0.30/0.20 & 0.19/0.15 & $-0.11$ \\
hiss & 7 & 0.83/0.71 & 0.56/1.00 & $-0.27$ \\
yawn & 20 & 0.09/0.05 & 0.39/0.30 & $+0.30$ \\
snore & 15 & 0.81/0.73 & 0.76/0.73 & $-0.05$ \\
\bottomrule
\end{tabular}
\end{table}

\subsection{Ablation of rebalancing strength}

\begin{figure}[!htbp]
    \centering
    \includegraphics[
        width=0.80\columnwidth
    ]{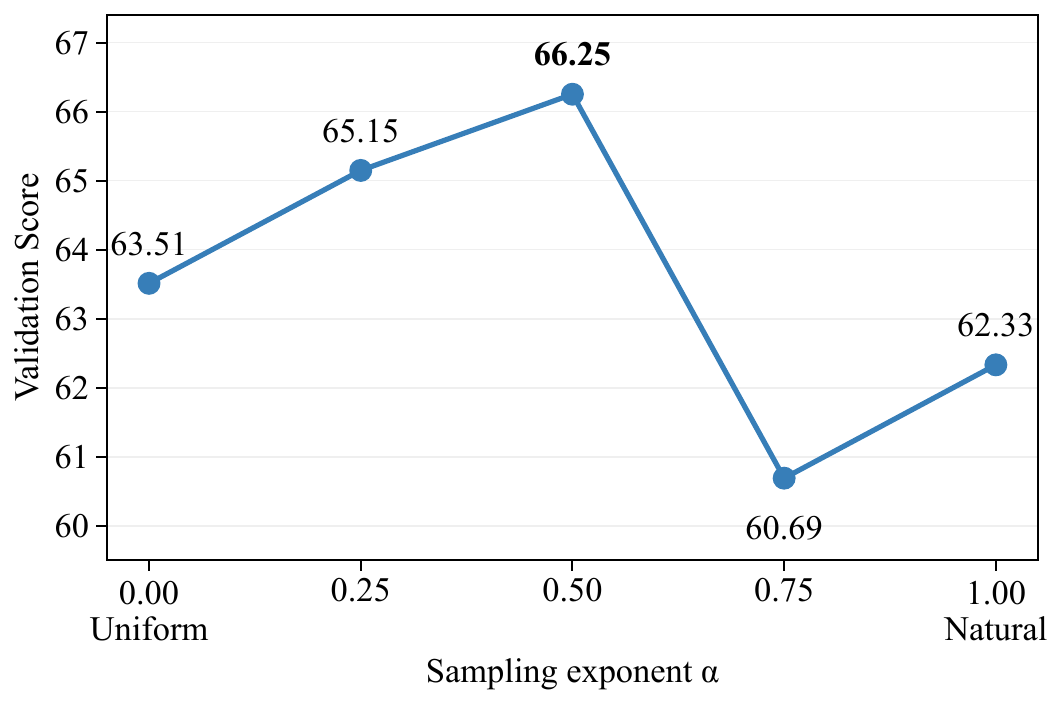}
    \setlength{\abovecaptionskip}{2pt}
    \caption{
        Local-validation score under different rebalancing strengths, indexed
        by $\alpha$, under the same backbone and optimization recipe.
    }
    \label{fig:alpha_ablation}
\end{figure}

To investigate how the strength of category rebalancing affects NVV-aware ASR performance on the local validation set, we compare the validation scores under different sampling exponents \(\alpha\), ranging from uniform category sampling (\(\alpha=0\)) to natural sampling (\(\alpha=1\)). As shown in Figure~\ref{fig:alpha_ablation}, the best score of 66.25 is achieved at \(\alpha=0.5\), whereas \(\alpha=0.75\) gives the lowest score of 60.69, suggesting that moderate rebalancing is more effective than either weaker or stronger rebalancing on our local validation set. As shown in Table~\ref{tab:local_validation}, compared with natural sampling, square-root sampling improves NVV recognition, while the English WER increases from 6.31\% to 7.10\%, indicating a modest trade-off between NVV recognition and lexical transcription performance. Meanwhile, uniform sampling (\(\alpha=0\)) is suboptimal, suggesting that retaining some original imbalance may be beneficial.

\subsection{Evaluation on MNV-17}
\label{subsec:mnv17}

For generalization, we evaluate the baseline, SQRT, and two-stage systems on 140 Track-1-compatible Mandarin utterances from MNV-17. As shown in Table~\ref{tab:mnv17}, the two-stage checkpoint achieves the lowest joint CER, while SQRT obtains higher NVV exact-match accuracy.

\begin{table}[!ht]
    \centering
    \caption{Results on the MNV-17 test split.}
    \label{tab:mnv17}
    \footnotesize
    \setlength{\tabcolsep}{2.5pt}
    \renewcommand{\arraystretch}{1.02}
    \begin{tabularx}{\linewidth}{@{}
        >{\centering\arraybackslash}X
        >{\centering\arraybackslash}p{0.22\linewidth}
        >{\centering\arraybackslash}p{0.30\linewidth}
        @{}}
        \toprule
        \textbf{System}
        & \textbf{Joint CER $\downarrow$}
        & \textbf{NVV accuracy $\uparrow$} \\
        \midrule
        Official Whisper baseline
        & 21.49\%
        & 25.00\% \\
        Qwen3-ASR 1.7B (SQRT)
        & 3.67\%
        & \textbf{63.57\%} \\
        Qwen3-ASR 1.7B (two-stage)
        & \textbf{3.21\%}
        & 60.71\% \\
        \bottomrule
    \end{tabularx}
\end{table}

\section{Conclusion}

Our data-centric NVV-aware ASR pipeline combines label harmonization,
power-based sampling, and uniform-category fine-tuning without changing the
backbone, objective, or output format. Square-root sampling performs best
on the local validation split among the tested single-stage settings, while
the two-stage system achieves our highest official submission score. On
MNV-17, the two-stage model yields lower joint CER, whereas SQRT gives higher
NVV exact-match accuracy. These results support category sampling as a
practical data-centric tool for NVV-aware ASR. For future work, we plan to
extend our approach to broader and more diverse datasets to further validate
its generalizability and effectiveness.

\clearpage
\bibliographystyle{IEEEtran}

\bibliography{mybib}

\end{document}